\documentclass[11pt]{article}
\usepackage[margin=1in]{geometry}
\usepackage[T1]{fontenc}
\usepackage{lmodern}
\usepackage{amsmath}
\usepackage{amssymb}
\usepackage{booktabs}
\usepackage{array}
\usepackage{tabularx}
\usepackage{listings}
\usepackage{url}
\usepackage{microtype}
\usepackage{enumitem}
\usepackage[hidelinks]{hyperref}

\newcommand{\uc}[2][1]{\makebox[#1\fontcharwd\font`\x][c]{#2}}

\newcolumntype{L}{>{\raggedright\arraybackslash}X}
\setlist[itemize]{leftmargin=1.5em}

\title{Modern Software Engineering with Verification Agents: A Spec-First Architecture for Autonomous Synthesis, Verification IRs, and Organizational Scale}
\author{Author: Jason Hickey\\
Affiliation: Independent Research\thanks{Disclaimer: The author is employed by Google Research. The views in this paper are of the author and do not necessarily reflect the views of his employer, or any employer.}\\
Email: \texttt{jasonh@gmail.com}}
\date{}

\begin{document}
\maketitle

\begin{abstract}
When autonomous coding agents synthesize implementation code, formal specifications, and mechanical proofs simultaneously, the software engineering hierarchy inverts: the formal specification becomes the primary human-reviewed artifact, while source code and proof scripts become disposable build outputs. This paper presents an end-to-end architecture for agent-authored, formally verified software engineering. First, we establish a logic-agnostic foundation combining a spec-first stack inversion, a zero-code-review workflow gated by automated continuous integration (CI) checks, and a minimal Verification Intermediate Representation (VIR) lowered via a thin, verified translator. Second, we pair an extensional Computational Type Theory (CTT) specification layer with Higher-Order Abstract Syntax (HOAS) and linear resource typing in the VIR---eliminating rewrite friction for synthesis agents while defining an experiment, its protocol frozen before the first reader, to measure whether removing boundary coercions improves human specification readability. Third, we scale collaboration across multi-team organizations using Personal Spec-Copilots and a Shared Synthesis Fleet that detects cross-team Semantic Merge Conflicts and fuses modular contracts into zero-copy binaries. We illustrate this architecture on a zero-copy authenticated message router and distinguish our existing empirical baselines (salt, SaltBench, x86lean, and MetaPRL) from newly proposed components.
\end{abstract}

\section*{1. Introduction}
For seventy years, much of the software stack---programming languages, optimizing compilers, software design patterns, and code review workflows---has been shaped around one scarce resource: human attention.

\begin{itemize}
\item \textbf{Programming Languages:} Mainstream languages introduced complex syntactic sugar, implicit type coercions, operator overloading, and intricate lifetime elision rules largely for the convenience of human readers and writers.
\item \textbf{Optimizing Compilers:} Production toolchains such as LLVM~\cite{lattner2004llvm} and GCC grew into large, unverified heuristic engines so that human engineers could write clean, readable, textbook code and still obtain high-performance machine binaries.
\item \textbf{Software Architecture:} Engineering practices enforced strict module indirection, deep object-oriented hierarchies, and ``Don't Repeat Yourself'' (DRY) boundaries so that a human developer could modify one subsystem without inadvertently breaking another.
\end{itemize}

With the rise of capable coding agents, this historical constraint has inverted: code generation is now cheap and abundant, and we take as our premise that human review, not generation, is now the bottleneck. Asking a senior software engineer (SWE) to manually audit 5,000 lines of agent-generated Rust or C++ for subtle memory bugs, race conditions, or security vulnerabilities is slow, exhausting, and error-prone.

However, when a coding agent generates a formal proof alongside its implementation---and a small, mathematically trusted proof kernel~\cite{barendregt2005challenge} mechanically verifies that the implementation satisfies a 25-line formal specification---the economics of software review change fundamentally. Human engineers no longer need to inspect how the implementation works or how the proof was constructed. Instead, engineering rigor concentrates on a much smaller, higher-leverage surface: verifying that the formal specification accurately captures intended system behavior.

\subsection*{1.1 The Five Architectural Pillars}
This paper proposes an end-to-end architecture structured around five pillars:

\begin{itemize}
\item \textbf{The Core Inversion (Spec as Source of Truth):} In greenfield development where AI agents synthesize implementation code, formal specifications, and mechanical proofs, the specification becomes the primary human-authored and human-reviewed artifact, while implementation code and proof scripts are treated as ephemeral build outputs regenerated on demand.
\item \textbf{Zero Human Code Review:} SWEs no longer read implementation diffs or proof scripts during change review; human attention is strictly conserved for module boundary specifications and automated operational telemetry (latency benchmarks, memory profiles, and automated vacuity and mutation gates).
\item \textbf{Shrinking the Compiler TCB via a Verification IR (VIR):} Once humans stop reading source code, languages no longer require human-ergonomic syntactic sugar or massive unverified optimizing compilers, converging instead toward a Verification IR (VIR) where agents perform aggressive optimizations directly in verified source and lower via a thin, CompCert-style verified translator~\cite{leroy2009compcert}.
\item \textbf{Designing for Readability and Rewrite Ergonomics (Extensional CTT + HOAS):} Pairing an extensional Computational Type Theory (CTT, pioneered by the Nuprl/MetaPRL family of provers~\cite{constable1986nuprl,hickey2003metaprl}) with Higher-Order Abstract Syntax (HOAS)~\cite{pfenning1988hoas} removes rewrite friction (``transport hell'' and binder-renaming boilerplate) for synthesis agents and removes coercions at module boundaries for human reviewers.
\item \textbf{Organizational Scaling via a Two-Tier Agent Fleet:} Across teams and open-source communities, SWEs collaborate on a modular specification graph using lightweight Personal Spec-Copilots backed by a Shared Synthesis Fleet that detects cross-team Semantic Merge Conflicts early and compiles modular contracts into globally fused machine code.
\end{itemize}

\subsection*{1.2 Goals and Explicit Non-Goals}
To turn source-level agentic verification into a practical engineering discipline at organizational scale, our architecture targets four primary goals:

\begin{itemize}
\item \textbf{Minimize Human Review Attention (Estimated >90\% Reduction):} Restrict human change-review surfaces strictly to public module boundary specifications and automated non-functional telemetry, eliminating manual line-by-line code review for verified modules. While the 90\% figure is an estimate, its first measurement is commissioned via the review-surface ratio (specification lines vs. implementation and proof lines) over the 181 conditions of record in SaltBench's pilot matrix, declared as a proxy for attention. The synthesis cost alongside this saving is already measured: on five Rust systems components refereed by a withheld test suite, asking the agent to specify and verify its own code cost more on all five, by a practical margin: no premium exceeded 2.8879\ensuremath{\times} and the three cheapest sat below 1.4\ensuremath{\times} (one model, three runs per condition), a bound of this population and not a promise about larger ones~\cite{hickey2026saltbench}.
\item \textbf{Shrink the Compiler Trusted Computing Base (TCB):} Move complex, bug-prone~\cite{yang2011csmith} program optimizations out of unverified compiler passes (such as LLVM) and into agent-generated, source-verified transformations over a dedicated VIR.
\item \textbf{Maximize Specification Readability and Agent Ergonomics:} Keep formal specifications mathematically clean, concise, and readable by generalist SWEs---free of proof witnesses, artificial termination fuel, or index-shifting plumbing---while minimizing proof-search friction for the agent.
\item \textbf{Scale Multi-SWE Collaboration:} Establish an organizational workflow where independent engineering teams compose modular specifications safely and detect logical contradictions---Semantic Merge Conflicts---before a single line of implementation code is synthesized.
\end{itemize}

Simultaneously, we explicitly rule out two common anti-patterns as non-goals:

\begin{itemize}
\item \textbf{Human-Editable or Stylistically ``Clean'' VIR:} The VIR is an intermediate synthesis target for agents, not a human programming language, and is never designed to be hand-edited, formatted, or stylistically reviewed by humans.
\item \textbf{Trusted Heuristics Inside the Verification Kernel:} LLM synthesis agents, proof-search heuristics, and optimization tactics remain 100\% untrusted; soundness relies exclusively on a minimal logical proof kernel~\cite{barendregt2005challenge} and a formally verified lowering pass.
\end{itemize}

\section*{2. Core Architecture: Specifications to Verification IR (Pillars 1--3)}
The first three pillars of the architecture are completely logic-agnostic and apply regardless of whether the underlying proof assistant is Lean~\cite{demoura2021lean4}, Rocq~\cite{bertot2004coqart}, Verus~\cite{lattuada2023verus}, Dafny~\cite{leino2010dafny}, or MetaPRL~\cite{hickey2003metaprl}. Removing human engineers from the source-code authoring loop triggers a structural inversion across the software stack.

\subsection*{2.1 The Stack Inversion (Pillar 1)}
Design practices that were essential when humans maintained source files become counterproductive overhead when agents synthesize and verify code directly from specifications.

\begin{center}\small
\begin{tabularx}{\linewidth}{@{}LLL@{}}
\toprule
\textbf{Stack Layer} & \textbf{Human-Centric Engineering} & \textbf{Spec-First Agentic Engineering} \tabularnewline
\midrule
Primary Artifact & Implementation source code (.cc, .rs, .go) & Modular formal specs and shared ontology \tabularnewline
Source Language / IR & Complex human-ergonomic languages & Minimal Verification IR (VIR) \tabularnewline
Optimization Layer & Unverified compiler passes (LLVM) & Agent-synthesized VIR rewrites \tabularnewline
Compiler Role & Heavyweight optimizer and codegen & Thin verified 1-to-1 lowering (CompCert-style) \tabularnewline
Modularity & Runtime indirection and interfaces & Spec-only; fused across modules in VIR \tabularnewline
Change Review & Line-by-line code diffs and unit tests & Boundary specs and CI vacuity/benchmark telemetry \tabularnewline
\bottomrule
\end{tabularx}
\end{center}

As summarized in the comparison above, traditional human-centric software engineering places implementation files (.cc, .rs, .go), heavy optimizing compilers, and runtime modularity boundaries at the center of the workflow, whereas Spec-First Agentic Engineering elevates modular formal specifications and a shared domain ontology to the primary artifact, delegates cross-module fusion and optimization to agent-generated VIR proofs, and reduces the compiler to a thin CompCert-style lowering layer~\cite{leroy2009compcert}.

\subsection*{2.2 The Zero-Code-Review Change Workflow (Pillar 2)}
If SWEs are forced to read both the formal specification and the implementation code, formal verification becomes a net tax on engineering velocity. In a Spec-First workflow, correctness is offloaded entirely to the mechanical proof kernel, and implementation code is treated like compiler-generated assembly---an opaque build artifact inspected only when profiling an unmodeled hardware bottleneck.

To prevent specification review from becoming a new cognitive bottleneck and to guard against an agent subtly gaming or weakening a specification, every change request is gated by four automated guardrails before a human reviewer opens it:

\begin{itemize}
\item \textbf{Boundary-Only Change Views:} Internal helper lemmas, loop invariants, and VIR implementation files are collapsed by default so the human SWE inspects only diffs to public module interface specifications and shared domain type definitions.
\item \textbf{Zero-Axiom and Spec Well-Formedness Gate:} A deterministic linter mechanically verifies that the proof relies on zero unverified axioms or escape hatches (sorry, admit, unsafe, assume), and proves that the specification itself is well-formed (\ensuremath{\vdash} Spec \ensuremath{\in} U\_i) so untyped ``junk terms'' cannot vacuously satisfy a contract.
\item \textbf{Automated Vacuity and Mutation Gates:} To ensure the agent has not made a theorem vacuously true~\cite{beer2001vacuity} via contradictory preconditions, CI requires an automatic witness proof (\ensuremath{\exists} x. Precond(x) \ensuremath{\wedge} Postcond(f(x))), while an adversarial mutation runner~\cite{demillo1978mutation} injects synthetic bugs into the generated VIR and verifies that the formal proof breaks on every mutant.
\item \textbf{Differential Litmus Vectors and Telemetry:} An independent critic agent generates concrete input/output test vectors directly from the natural-language prompt and proves the specification implies them, alongside automated operational telemetry covering latency, heap allocations, and binary size.
\end{itemize}

\subsection*{2.3 Target Language Evolution: The Verification IR (Pillar 3)}
When verification is performed at the source level, both the formal semantics of the source language and the correctness of the compiler sit inside the TCB. Mainstream languages (C++, Rust, Go) have massive formal semantics because they optimize for human ergonomics, while production compilers (LLVM, GCC) contain large bodies of unverified heuristic code that regularly suffer from miscompilation bugs~\cite{yang2011csmith}. Once humans stop reading or editing source code, three simplifications follow:

\begin{itemize}
\item \textbf{Convergence to a Verification IR (VIR):} Syntactic sugar becomes a pure liability, so the target language evolves into a minimal VIR with a tiny, total formal semantics (a short specification rather than a standard of well over a thousand pages) where all types, memory effects, and control-flow transitions are 100\% explicit.
\item \textbf{Moving Optimizations from the Compiler to Agent VIR:} Instead of trusting an unverified LLVM pass to unroll loops, inline closures, or fuse passes across modules, the coding agent writes already-unrolled, cache-aligned, specialized code directly in the VIR and proves it satisfies the high-level specification. Because the VIR is already optimized, the underlying compiler shrinks into a thin, formally verified 1-to-1 lowering translator (CompCert~\cite{leroy2009compcert} or CakeML~\cite{kumar2014cakeml} style) targeting a mechanized ISA semantics such as x86lean~\cite{x86lean2026} (a Lean 4 x86-64 user-level semantics built from public sources and differentially validated against the ACL2 x86isa model, with the rows it cannot carry checked on real processors).
\item \textbf{Eliminating Source-Level ``Clean Code'' Overhead:} Traditional rules like DRY~\cite{hunt1999pragmatic} and deep interface indirection create severe aliasing and framing headaches for formal verification; in our model, DRY lives strictly in the specification while the agent generates flat, aggressively inlined VIR code.
\end{itemize}

\subsection*{2.4 Memory Management Without Garbage Collectors or Borrow Checkers}
Removing human engineers from the source layer also resolves the classic systems programming trilemma of memory management: manual malloc/free (fast but memory-unsafe), tracing garbage collection~\cite{jones2023gchandbook} (safe for humans, but adds a massive runtime to the TCB and can threaten tail-latency bounds), and lexical borrow checking~\cite{jung2018rustbelt} (safe and fast, but forces a conservative static heuristic that rejects some aliasing patterns that are safe).

\begin{center}\footnotesize
\begin{tabularx}{\linewidth}{@{}LLLL@{}}
\toprule
\textbf{Memory Model} & \textbf{TCB and Cost Impact} & \textbf{Agent VIR Synthesis Impact} & \textbf{Architectural Role} \tabularnewline
\midrule
Tracing GC (Go / Java / OCaml) & Poor; large runtime TCB, breaks tail latency & Easy; no deallocation proofs & Avoid for systems code \tabularnewline
Lexical Borrow Checker (Rust) & Good TCB; rigid expressiveness & High friction against heuristics & Replaced by proof capabilities \tabularnewline
Linear Capabilities (Perceus / FBIP) & Zero runtime TCB; exact byte accounting & Natural local equational rewrite & Primary mechanism (most code) \tabularnewline
Lexical Regions (Arena / Verus) & Zero runtime TCB; O(1) bulk free & High leverage; no node lifetimes & Graphs and request scopes (the remainder) \tabularnewline
\bottomrule
\end{tabularx}
\end{center}

As shown in the table above, human-authored specifications remain purely functional (using value semantics with optional bulk memory bounds in the Cost Spec, such as maxHeap(f) \ensuremath{\le} 4 KB), while the agent manages physical memory in the VIR primarily through Perceus/FBIP linear capabilities~\cite{reinking2021perceus,lorenzen2023fp2} (we expect most code: trees, buffers, streaming pipelines) and Arena/Verus-style lexical regions~\cite{tofte1997region,lattuada2023verus} with separation tokens (the remainder: cyclic graphs, request scopes).

\subsection*{2.5 Performance Engineering and Implementation Directives}
To allow human engineers to steer concrete data-structure choices and guarantee high performance without polluting functional specifications, the architecture separates correctness, performance bounds, and implementation guidance into three orthogonal channels.

\begin{center}\footnotesize
\begin{tabularx}{\linewidth}{@{}LLLL@{}}
\toprule
\textbf{Channel} & \textbf{Human SWE Artifact} & \textbf{Trusted Contract} & \textbf{Enforcement Mechanism} \tabularnewline
\midrule
1. Functional Spec & Pure math over abstract types: Map(K, V), lookup(insert(m, k, v), k) = v & Yes (Hard Gate) & Proof kernel checks extensional equality \tabularnewline
2. Cost and Resource Spec & Asymptotic or concrete bounds: steps(lookup) \ensuremath{\le} 4\ensuremath{\cdot}log\ensuremath{_{2}}(N), heapAlloc = 0 & Yes (Hard Gate) & Kernel ANF step weights + CI telemetry \tabularnewline
3. Implementation Directives & 1-line strategy hint: \#[realize(Index, via = BTree(B = 16))] & No (Untrusted) & Steers agent tactic search and seeding \tabularnewline
\bottomrule
\end{tabularx}
\end{center}

This three-channel separation supports two synergistic workflows:

\begin{itemize}
\item \textbf{Non-Binding Implementation Directives and Quotient Realizations:} In Extensional CTT, an abstract specification type like Map(K, V) is defined as a Quotient / Partial Equivalence Relation (PER) over concrete realizations~\cite{constable1986nuprl}. When an SWE attaches a directive such as \#[realize(Index, via = BTree(B = 16))] (or fanout = 16), downstream modules and security reviewers still see only the pure mathematical Map(K, V), and swapping the directive to RadixTree breaks zero downstream specifications or caller theorems.
\item \textbf{Provable Cost Semantics and Agentic Pareto Search:} Because the VIR is in A-Normal Form (Functional SSA)~\cite{flanagan1993anf,appel1998ssa}, every step and allocation has an explicit cost weight checked against bounds like steps(lookup) \ensuremath{\le} 4\ensuremath{\cdot}log\ensuremath{_{2}}(N) and heapAlloc = 0. Meanwhile, micro-architectural tuning (cache-line alignment, SIMD lane width, prefetching) is handled by the Shared Synthesis Fleet as an autonomous Pareto search problem across candidate realizations benchmarked against production replay traces.
\end{itemize}

\section*{3. Readability and Rewrite Ergonomics: Extensional Logic and HOAS (Pillar 4)}
Putting the Spec \ensuremath{\rightarrow} VIR architecture into practice exposes two critical bottlenecks: Specification Readability for the human reviewer and Rewrite Ergonomics for the synthesis agent. We address these through (1) an Extensional Logical Framework based on Computational Type Theory (CTT---underlying Nuprl~\cite{constable1986nuprl}, MetaPRL~\cite{hickey2003metaprl}, and RedPRL~\cite{angiuli2018redprl}) and (2) Higher-Order Abstract Syntax (HOAS)~\cite{pfenning1988hoas,chlipala2008phoas} in the VIR.

Stated at the strength supported by current evidence, CTT buys the synthesis agent frictionless rewriting (a result measured for a hand-written formal compiler, Section 3.2, which we expect to carry over to synthesis agents), and for the human reader it removes coercions at module boundaries while shifting the well-formedness guarantee from the typechecker to a CI gate (Section 2.2, Gate 2). Whether that tradeoff yields a net improvement in human specification readability is an open empirical question addressed by the experiment in Section 3.3, whose protocol is frozen before the first reader is run.

\subsection*{3.1 Extensional CTT vs. Intensional CIC/Lean}
Intensional dependent type theories with strong normalization---such as the Calculus of Inductive Constructions (CIC)~\cite{coquand1988coc,bertot2004coqart} in Lean~\cite{demoura2021lean4} and Rocq---are the basis of the most widely used provers, and decidable type checking of proof terms is one of their design goals. However, strong normalization and intensional typing require termination arguments for potentially diverging loops or partial parsers, and represent subset constraints as pairs of data and proof that cross module boundaries via explicit coercions. In an agent-driven workflow where humans never read proof scripts, Computational Type Theory (CTT) layers semantic, extensional types~\cite{martinlof1984itt} over untyped terms without requiring strong normalization~\cite{constable1986nuprl}.

\begin{center}\footnotesize
\begin{tabularx}{\linewidth}{@{}LLLL@{}}
\toprule
\textbf{Framework Dimension} & \textbf{Extensional CTT Style} & \textbf{CIC / Lean / Rocq Style} & \textbf{Impact on Readability and Search} \tabularnewline
\midrule
Consistency Mechanism & Semantic PERs over untyped terms & Syntactic strong normalization & CTT specifies partial programs directly \tabularnewline
Subset and Quotient Types & Extensional and erased \{ x : A \ensuremath{\mid} P(x) \} & Intensional Sigma pairs \ensuremath{\langle}x, h\_P\ensuremath{\rangle} & CTT needs no boundary coercions (\ensuremath{\uparrow}b, .val) \tabularnewline
Rewriting Under Binders & Frictionless untyped lam\{x. e\} & Typed binders require Eq.rec / HEq & CTT removes transport casts in search (not measured here) \tabularnewline
HOAS Support & Native primitive lam\{x. e\} & Blocked by strict positivity & Direct meta-level \ensuremath{\beta}-reduction in CTT \tabularnewline
Typechecking Decidability & Undecidable; proved via tactics & Decidable automatic kernel check & Agents discharge \ensuremath{\vdash} Spec \ensuremath{\in} U\_i in CI \tabularnewline
\bottomrule
\end{tabularx}
\end{center}

Two foundational technical considerations govern the use of CTT in this architecture:

\begin{itemize}
\item \textbf{Predicativity in CTT:} Because CTT's universe hierarchy is predicative (U\ensuremath{_{1}} \ensuremath{\subset} U\ensuremath{_{2}} \ensuremath{\subset} U\ensuremath{_{3}} ...), quantifying over types or state predicates bumps a specification to the next universe level. However, because CTT terms are untyped and recursive types (\ensuremath{\mu}) plus dependent intersections (x:A \ensuremath{\cap} B[x])~\cite{kopylov2003depint} are built directly into the PER semantics, implementation code remains universe-free, and schematic universe polymorphism infers level offsets invisibly in the background.
\item \textbf{The Trusted Core in CTT:} While intensional systems like Lean already enjoy small, independently re-implemented proof kernels~\cite{carneiro2024lean4lean}, CTT historically placed its refiner in the trusted base. In our agentic architecture, the refiner, every tactic, and the synthesis agent are all untrusted: a recorded proof certificate (as supported by MetaPRL) is checked against a small core whose rules are verified against the PER semantics, building on Anand and Rahli's Coq formalization of Nuprl's semantics (ITP 2014)~\cite{anand2014towards}. A verified certificate checker for the VIR's logic is a proposed component (Section 7), and Pillar 3's TCB claim is conditional on it.
\end{itemize}

\subsection*{3.2 Eliminating Agent Rewrite Friction: HOAS vs. First-Order Binders}
Because the coding agent performs aggressive source-level optimizations---inlining functions across modules, specializing closures, and fusing loops---every friction point in variable binding and substitution multiplies token consumption and proof-search depth. In empirical formal compiler work (Hickey and Nogin, Formal Compiler Construction in a Logical Framework~\cite{hickey2006formal,hickey2003formaltr}), representing binders with HOAS in MetaPRL reduced the formal rewrite rules for CPS conversion, closure conversion, and code generation to 484 lines of formal mathematics---compared to 12,000 lines in a traditional compiler---and cut implementation time by 3\ensuremath{\times}.

\begin{center}\footnotesize
\begin{tabularx}{\linewidth}{@{}LLLL@{}}
\toprule
\textbf{VIR Requirement} & \textbf{HOAS / PHOAS Representation} & \textbf{CompCert First-Order (IDs + Env)} & \textbf{Architectural Verdict} \tabularnewline
\midrule
Agent Inlining & Free via meta-level \ensuremath{\beta}-reduction & High proof tax; fresh ID maps & HOAS wins for optimization throughput \tabularnewline
Code Motion and Fusion & Compact local capture-avoiding rewrites & High friction; environment shifting lemmas & HOAS wins on token economy \tabularnewline
Operational Cost Model & Hard in pure HOAS; hides steps & Natural explicit state machine & First-order wins at machine boundary \tabularnewline
\bottomrule
\end{tabularx}
\end{center}

\subsubsection*{The Hybrid VIR Solution: Three-Stage Lowering Pipeline}
To combine the algebraic rewrite speed of HOAS with the cycle-accurate operational semantics of CompCert~\cite{leroy2009compcert} at the hardware boundary, our VIR uses a three-stage lowering pipeline:

\begin{itemize}
\item \textbf{Linear Functional SSA with Untyped HOAS Binders:} Every intermediate computation in the VIR is explicitly let-bound in A-Normal Form---written in display form as let\_lin\{x = op(a, b) in e\} and desugaring to the canonical MetaPRL operator let\_lin\{op(a, b); x. e[x]\}---which is mathematically isomorphic~\cite{appel1998ssa} to Static Single Assignment (SSA)~\cite{cytron1991ssa} with block arguments~\cite{lattner2021mlir}, making operational execution steps and allocation costs 100\% explicit while providing free capture-avoiding substitution~\cite{pfenning1988hoas,miller1991patterns}.
\item \textbf{Linear and Affine Resource Typing:} Combining HOAS binders with linear ownership types~\cite{girard1987linear} ensures that a functional binding let\_lin\{r1 = add(r1, r2) in e\} statically guarantees the previous state of r1 is never referenced again, compiling directly to an in-place destructive hardware register or buffer update without continuation-passing overhead.
\item \textbf{Register Allocation as Verified \ensuremath{\alpha}-Conversion:} Rather than verifying a complex graph-coloring allocator~\cite{chaitin1982regalloc} inside the TCB, an untrusted agent heuristic proposes a physical register assignment that the logical framework validates~\cite{pnueli1998tv,rideau2010regalloc} as a scope-preserving, non-aliasing \ensuremath{\alpha}-conversion before emitting straight-line machine code via a thin CompCert-style assembler.
\end{itemize}

\subsubsection*{HOAS-Native Memory Management}
Pairing CTT with HOAS turns the memory model of Section 2.4 into four local syntactic operations where variable scoping enforces memory safety:

\begin{itemize}
\item \textbf{Functional-But-In-Place (FBIP) Reuse via HOAS Rewrites:} Pairing linear bindings (let\_lin) with Perceus-style reuse rewrites~\cite{reinking2021perceus,lorenzen2023fp2} (free(x); let\_lin\{y = alloc(sz) in e[y]\} \ensuremath{\leftrightarrow} let\_lin\{y = reuse(x) in e[y]\}) keeps the specification pure while mutating memory in place with zero heap allocation.
\item \textbf{Lexical Regions as HOAS Binders (Free Escape Analysis):} The VIR provides a region binder let\_region\{\ensuremath{\rho}. e[\ensuremath{\rho}]\} supporting O(1) bump allocation (alloc\_in(\ensuremath{\rho}, sz)); because \ensuremath{\rho} is a bound meta-variable in HOAS, the standard typing rule requires \ensuremath{\rho} \ensuremath{\notin} FV(ResultType), automatically guaranteeing~\cite{tofte1997region,launchbury1994st} that no pointer tied to region \ensuremath{\rho} can escape and enabling O(1) bulk deallocation on scope exit.
\item \textbf{First-Class Permission Tokens for Intrusive Aliasing:} For kernel page tables, doubly-linked lists, or lock-free ring buffers, CTT separates the copyable integer address (addr : U64) from the linear ownership predicate (perm : Own(addr, T))~\cite{reynolds2002sl,lattuada2023verus} threaded through let\_lin binders.
\item \textbf{Concurrency and Shared-Memory Parallelism via Fractional Permissions:} Following Concurrent Separation Logic~\cite{ohearn2007resources} (Iris~\cite{jung2018iris} / Verus~\cite{lattuada2023verus} style), an exclusive token Own(addr, T) splits via a verified HOAS rewrite into fractional~\cite{boyland2003fractional} read-only tokens Read(addr, T, q/2) across concurrent readers and reassembles upon join, while channel sends (send\{ch, addr, perm; e\}) linearly consume perm in the sender's scope and transfer it to the receiver.
\end{itemize}

\subsection*{3.3 A Planned Empirical Evaluation of Specification Readability}
While the agent-side rewrite gains in Section 3.2 rest on published measurements of a hand-written formal compiler, the human- and model-reader gain has not yet been measured and carries a natural confound: language models and engineers have very likely read far more Lean than CTT.

To separate intrinsic readability from familiarity, we define an experiment whose protocol is drafted, and frozen before any reader is evaluated:

\begin{itemize}
\item \textbf{Paired Specifications:} Five specifications---(1) the router ACL, (2) a wire-format round trip, (3) a token-bucket cost bound, (4) an abstract Map with a quotient realization, and (5) a region-escape property---are each written in their best idiomatic style in both Lean and CTT and verified by an independent third party to state the exact same property.
\item \textbf{Familiarity Baseline:} A third rendering of each specification is prepared in neutral mathematical notation as a familiarity baseline.
\item \textbf{Evaluation Protocol:} Comprehension questions are authored from plain-English intent by a non-author, including one trap question per pair that the specification does not answer, alongside a certificate control (a specification with one deliberately wrong clause to catch readers answering from memory rather than the page).
\item \textbf{Reader Cohorts:} Fresh model readers from two LLM vendors and a human cohort of software engineers unfamiliar with both systems are scored strictly by artifact with raw outputs preserved.
\item \textbf{Declared Asymmetry:} The Lean specifications are mechanically checked by Lean, whereas the CTT specifications are checked by hand because none of the CTT provers we know of (Nuprl, MetaPRL, RedPRL) has a toolchain we could use for this experiment---an asymmetry that directly informs the well-formedness gate requirement of Section 2.2.
\end{itemize}

\section*{4. Organizational Scaling: Multi-SWE Collaboration (Pillar 5)}
Treating verification agents as isolated ``one-agent-per-human'' tools across an organization of hundreds of SWEs would be analogous to asking every developer to run a private compiler and check raw binaries into Git. Instead, organizational scaling relies on a Two-Tier Agent Infrastructure and new collaboration primitives.

\subsection*{4.1 Two-Tier Agent Infrastructure}
\begin{itemize}
\item \textbf{Tier 1 --- Personal Spec-Copilot (1 per SWE):} Each engineer interacts with a low-latency Spec-Copilot in their local IDE or change-review interface that helps formalize domain intent into clean CTT specifications, translates formal contracts back into plain-English edge-case summaries, queries upstream and downstream team contracts, and runs instant vacuity checks.
\item \textbf{Tier 2 --- Shared Synthesis Fleet (Organization Compute Backplane):} Operating as continuous shared infrastructure, the Shared Synthesis Fleet consumes the repository's global graph of approved specifications, synthesizes cross-module HOAS VIR, maintains proofs across repository-wide spec updates, checks cross-team logical consistency, and compiles globally fused machine binaries.
\end{itemize}

\subsection*{4.2 Collaboration Primitives: Semantic Merges, Ontology, and TCB Debt}
\begin{itemize}
\item \textbf{Semantic Merge Conflicts Replace Textual Git Conflicts:} When Engineer A and Engineer B submit concurrent specification changes (S\_A and S\_B), the Shared Fleet searches within a compute budget for the vacuity witness required by S\_A \ensuremath{\wedge} S\_B (Section 2.2, Gate 3). Because satisfiability is undecidable in general~\cite{church1936note}, the primitive is the witness rather than a decision procedure: if found, the fleet synthesizes a unified VIR satisfying S\_A \ensuremath{\cap} S\_B; if not found within budget, the fleet halts the merge and issues a Semantic Merge Conflict containing either a concrete counterexample state (when a refutation is found) or the unsatisfied conjunction (when the search exhausts its budget).
\item \textbf{Decoupling Conway's Law~\cite{conway1968committees} from Runtime Performance:} Five teams can own five cleanly separated modular specifications while the Shared Synthesis Fleet fuses them via HOAS \ensuremath{\beta}-reduction into a single, zero-copy VIR pipeline without microservice or data-copying boundaries.
\item \textbf{Ontology Governance and Open-Source ``Spec Commons'':} Senior and Staff SWEs shift their coordination effort from reviewing code style to governing the Shared Domain Ontology---the minimal set of base types and predicates that allow independent team specs to compose without adapter boilerplate. Open-source communities maintain a shared reference specification and accept zero-trust VIR + Proof Certificate contributions (such as hardware-specific SIMD optimizations) whenever the kernel verifies the proof in milliseconds and benchmarks confirm a Pareto improvement.
\item \textbf{Proof-Timeout Escalation and Tracked TCB Debt:} When a valid specification exceeds the Shared Synthesis Fleet's proof compute budget, the fleet follows a two-step escalation protocol rather than asking humans to debug VIR proof scripts: (1) it prompts the SWE via their Personal Copilot to add a 1-line Strategy Directive (Section 2.5) or factor the monolithic spec into two smaller module contracts; (2) for urgent production hotfixes, it isolates the unproven subgoal behind a dynamic Runtime Contract Monitor (assert!)~\cite{meyer1992dbc}, logging the assumption explicitly in the repository's TCB Debt Ledger until background proof search closes the obligation.
\end{itemize}

\subsection*{4.3 Brownfield Adoption Regimes}
Real-world adoption must coexist with massive brownfield C++ and Rust codebases, separating subsystems into two distinct regimes based on coupling and behavioral surface.

\begin{center}\small
\begin{tabularx}{\linewidth}{@{}LLL@{}}
\toprule
\textbf{Brownfield Regime} & \textbf{Architectural Characteristics} & \textbf{Recommended Migration Path} \tabularnewline
\midrule
Regime A: Bounded-Surface / High-Risk & Wire parsers, crypto/TPM, ACLs, narrow VMM device models & Spec extraction, differential fuzzing against legacy oracle, VIR re-synthesis behind C-ABI shim \tabularnewline
Regime B: Deeply Coupled Ecosystem & Large stateful C++ object graphs and bidirectional C++/Rust interop & Stay on host platform (Safe Rust + Crubit~\cite{crubit}) and apply host-language verification (Verus~\cite{lattuada2023verus}) \tabularnewline
\bottomrule
\end{tabularx}
\end{center}

In Regime A, teams avoid verifying messy legacy C++ in place by using the legacy binary as a behavioral oracle for adversarial differential fuzzing during spec extraction, locking the human-audited CTT spec, and replacing the legacy implementation behind a C-ABI shim guarded by runtime boundary monitors.

\section*{5. Case Study 1 (Technical): Zero-Copy Authenticated Message Router}
To demonstrate Pillars 1--4 in action, we examine the specification, synthesis, and verification of a Zero-Copy Authenticated Message Router that ingests raw network frames from a ring buffer, parses the wire header and topic identifier, verifies sender permissions against an Access Control List (ACL), enforces a token-bucket rate limit, and forwards authorized payload slices downstream.

\subsection*{5.1 The Engineering Dilemma and Modular Specifications}
In traditional systems, separating wire parsing, security ACLs, and rate-limiting into clean modules requires intermediate allocations (ParsedFrame) and slice copies across interfaces, whereas hand-fusing them into a single C/Rust state machine creates an inscrutable tangle of pointer arithmetic where an off-by-one error risks remote code execution. Under the Spec-First Agentic approach, teams establish a 10-line Shared Ontology (WireBuf, Frame, Principal, Topic) and author three independent specifications.

\begin{center}\small
\begin{tabularx}{\linewidth}{@{}LLL@{}}
\toprule
\textbf{Specification} & \textbf{Mathematical Contract (CTT Syntax)} & \textbf{Engineering Guarantee} \tabularnewline
\midrule
Spec A (Wire Protocol) & \ensuremath{\forall} f : Frame. parse(encode(f)) = inl(f) & Round-trip framing and serialization correctness \tabularnewline
Spec B (Security ACL) & \ensuremath{\forall} b : ValidFrame, p : Principal. route(p, b) = Allow(s) \ensuremath{\Rightarrow} frame\_topic(b) \ensuremath{\in} ACL(p) & Unauthorized topics are never routed downstream \tabularnewline
Spec C (Rate-Limit and Cost) & \ensuremath{\forall} b : ValidFrame, p : Principal. (route(p, b) = Allow(s) \ensuremath{\Rightarrow} tokens(p) > 0) \ensuremath{\wedge} heapAlloc = 0 \ensuremath{\wedge} steps \ensuremath{\le} 32 + 2\ensuremath{\cdot}len(b) & Token-bucket enforcement, zero heap allocation, O(len(b)) latency \tabularnewline
\bottomrule
\end{tabularx}
\end{center}

\subsubsection*{Contrast Exhibit: Idiomatic Lean vs. Extensional CTT}
Comparing each logic in its best idiomatic style illustrates the boundary distinction analyzed in Section 3.1.

\par\medskip\noindent\textbf{Exhibit A.1 --- Lean Specification (idiomatic Prop-carrying subtype, total accessors):}\par\nopagebreak
\begin{lstlisting}
abbrev WireBuf := ByteArray
def readU32 (b : WireBuf) (i : Nat) : UInt32 := `\uc{$\ldots$}`   -- total; exact on a ValidFrame

def ValidFrame := { b : WireBuf // 24 `\uc{$\le$}` b.size `\uc{$\wedge$}` readU32 b 0 = MAGIC }
def frameTopic (b : ValidFrame) : TopicId := readU32 b 8

theorem route_acl_safe (b : ValidFrame) (p : Principal) :
    IsRouted (route p b) `\uc{$\rightarrow$}` CanAccess p (frameTopic b)
\end{lstlisting}

\par\medskip\noindent\textbf{Exhibit A.2 --- CTT Specification (Extensional Subset Types --- Zero Clutter):}\par\nopagebreak
\begin{lstlisting}
`\uc{$\forall$}` b : ValidFrame, p : Principal.
  IsRouted(route(p, b)) `\uc{$\Rightarrow$}` CanAccess(p, frame_topic(b))
\end{lstlisting}
Both statements share the same top-level shape: fuel is absent from the Lean version because route is defined by well-founded recursion at the definition, and the proof witness is erased in Prop. The remaining difference lies at the module boundary: a CTT ValidFrame is a WireBuf, whereas a Lean ValidFrame is a pair whose value crosses boundaries through a coercion (\ensuremath{\uparrow}b, .val) visible in goals and error messages, while CTT shifts statement well-formedness to CI Gate 2.

\subsection*{5.2 Agent VIR Synthesis and HOAS Loop Fusion}
Starting from the conjunction Spec A \ensuremath{\cap} Spec B \ensuremath{\cap} Spec C, a naive compositional implementation executes three sequential passes over the packet buffer:

\par\medskip\noindent\textbf{Exhibit B.1 --- Unfused Compositional VIR Term:}\par\nopagebreak
\begin{lstlisting}
let_lin{ (b1, hdr) = parse_header(b0) in
let_lin{ (b2, tok) = check_rate_limit(b1, hdr.sender) in
let_lin{ (b3, ok)  = verify_acl(b2, hdr.sender, hdr.topic) in
  emit_payload(b3, hdr.payload_offset) }}}
\end{lstlisting}
Directly under the let\_lin HOAS binders, the synthesis agent inlines verify\_acl and check\_rate\_limit via meta-level \ensuremath{\beta}-reduction and applies the algebraic slice-coalescing rewrite advance(advance(b, i), j) \ensuremath{\leftrightarrow} advance(b, i + j) (subject to i \ensuremath{\ge} 0 \ensuremath{\wedge} j \ensuremath{\ge} 0 \ensuremath{\wedge} i + j \ensuremath{\le} len(b)). Because CTT binders are untyped, rewriting inside let\_lin\{b2 = ...\} folds all separate header and topic bounds checks into a single upfront guard (len(b0) \ensuremath{\ge} 24) and eliminates the intermediate hdr structure without requiring dependent HEq transport casts across length-indexed slice types (Slice (n - 4 - k) to Slice (n - (4 + k))).

\subsection*{5.3 The Zero-Code-Review Gate Scorecard}
When synthesis completes, the 350 lines of bit-twiddling VIR and 1,200 lines of tactic proof scripts are collapsed by default, and the human reviewer inspects only the 15-line specification diff alongside the automated change scorecard (illustrative figures showing gate structure).

\begin{center}\small
\begin{tabularx}{\linewidth}{@{}LLL@{}}
\toprule
\textbf{Automated Gate} & \textbf{Mechanical Check Performed} & \textbf{Status} \tabularnewline
\midrule
1. Kernel Proof and Well-Formedness & Verifies \ensuremath{\vdash} route\_impl \ensuremath{\in} Spec\_A \ensuremath{\cap} Spec\_B \ensuremath{\cap} Spec\_C with 0 axioms/admits & PASS (42 ms) \tabularnewline
2. Vacuity and Reachability Gate & Proves witness \ensuremath{\exists} p, b. IsRouted(route(p, b)) & PASS \tabularnewline
3. Adversarial Spec Mutation & Injects 10 synthetic bugs into VIR and confirms proof fails on all 10 & PASS (10/10 killed) \tabularnewline
4. Concrete Litmus Vectors & Evaluates 5 independent concrete I/O vectors from natural-language prompt & PASS (5/5 matched) \tabularnewline
5. Operational Telemetry & Lowered x86-64 profile: 0 heap allocs, 18 cycles/pkt (P99), 1.4 KB binary & PASS \tabularnewline
\bottomrule
\end{tabularx}
\end{center}

As shown in the scorecard above, every mechanical, vacuity, mutation, litmus, and operational check passes automatically before human review begins.

\section*{6. Case Study 2 (Organizational): Multi-Team Collaboration}
To illustrate Pillar 5 across a live organization, we follow three engineers collaborating on the Message Router---Alice (Wire Protocol Team), Bob (Security Team), and Carol (SRE / Performance Team)---across four scenarios. The scenarios and their figures are illustrative.

\begin{center}\small
\begin{tabularx}{\linewidth}{@{}LLL@{}}
\toprule
\textbf{Scenario} & \textbf{Traditional Software Engineering} & \textbf{Spec-First Agentic Organization} \tabularnewline
\midrule
Act I: Cross-Boundary Feature (8-byte CRC64 header) & Breaks auth offsets and zero-copy logic; 3-team migration and 1,500-line review & Alice edits 12 lines in Spec A; Bob's ValidFrame spec needs 0 edits; fleet re-fuses VIR \tabularnewline
Act II: Flawed Spec (Contradictory quarantine rule) & Unit tests mock setup and pass; dead security rule ships to production & Bob's Copilot Vacuity Gate fails \ensuremath{\exists} b. QuarantinePrecond(b) in 3 seconds \tabularnewline
Act III: Semantic Merge Conflict (64B log vs. 0B DDoS cap) & Separate files (auth.rs, limits.rs) merge cleanly in Git; OOM crash in prod & Fleet finds refutation \ensuremath{\vdash} False when rate > 10M \ensuremath{\wedge} drop(b) and blocks merge \tabularnewline
Act IV: Zero-Trust Open-Source AVX-512 SIMD PR & Maintainers hesitate to merge 2,000 lines of intrinsics due to backdoor risk & Kernel verifies certificate in 40 ms; CI confirms 3.2\ensuremath{\times} gain; merges with 0 code review \tabularnewline
\bottomrule
\end{tabularx}
\end{center}

Across all four acts, human attention remains bounded to domain-level intent and cross-team ontology contracts while the Shared Synthesis Fleet absorbs cross-module fusion, proof maintenance, and regression verification.

\section*{7. Risks, Roadmap, Related Work, and Implementation Status}

\subsection*{7.1 Risks, Limitations, and Mitigations}
Shifting human oversight from implementation code to formal specifications introduces three primary failure modes governed by architectural and organizational controls:

\begin{itemize}
\item \textbf{Specification Gaming (``Reward Hacking'') [High Severity]:} An agent might subtly weaken a specification precondition to make a difficult proof succeed; this is mitigated by locking specification files as read-only during VIR synthesis and enforcing automated Vacuity Proofs and Adversarial Spec Mutation Testing in CI before human review.
\item \textbf{Unmodeled Boundary Semantics (FFI / Hardware) [Medium Severity]:} Bugs in raw OS syscalls, DMA controllers, or hardware errata outside the formal model are mitigated by isolating raw I/O inside a tiny, human-audited Hardware Abstraction Layer (HAL) paired with runtime contract monitors in production.
\item \textbf{Ontology Fragmentation [Medium Severity]:} Incompatible base types across teams requiring brittle adapter proofs are prevented by elevating Ontology Architecture to a primary Staff/Principal SWE responsibility.
\end{itemize}

\subsection*{7.2 Phased Adoption Roadmap}
Organizations can adopt this architecture incrementally across three phases without a flag-day rewrite:

\begin{itemize}
\item \textbf{Phase 1 --- Spec-First Review and Differential Oracles on Host Languages (Immediate):} Deploy Personal Spec-Copilots and automated vacuity/mutation gates over existing C++ and Rust workflows (e.g., Verus on Rust components and differential fuzzing on bounded C++ subsystems) while tracking Verified Coverage.
\item \textbf{Phase 2 --- CTT + HOAS VIR Pilot on High-Assurance Bounded Subsystems (Near-Term):} Target isolated Regime A components (wire parsers, crypto/TPM routines, ACL evaluators, narrow VMM device models) for full Spec-Extraction and HOAS VIR synthesis behind verified C-ABI shims.
\item \textbf{Phase 3 --- Shared Synthesis Fleet and Cross-Module Fusion at Scale (Long-Term):} Connect modular team specifications into a repository-wide Spec Graph governed by a Shared Domain Ontology to detect Semantic Merge Conflicts and fuse adjacent VIR modules across team boundaries.
\end{itemize}

\subsection*{7.3 Related Work}
This architecture sits at the intersection of six research lineages and differs from each in a specific dimension:

\begin{itemize}
\item \textbf{Verification-Aware Languages:} Dafny~\cite{leino2010dafny}, F*~\cite{swamy2016fstar}, and Verus~\cite{lattuada2023verus} place specifications beside human-written code, and Aeneas~\cite{ho2022aeneas} translates Rust into Lean for proof; Pillar 1 removes the human as implementation author, and Pillar 3 removes the human-ergonomic source language.
\item \textbf{Verified Compilers:} CompCert~\cite{leroy2009compcert} and CakeML~\cite{kumar2014cakeml} verify optimizing compiler passes inside the TCB; Pillar 3 moves optimizations above the compiler into agent-verified VIR rewrites, retaining only a thin verified lowering layer.
\item \textbf{Computational Type Theory:} Nuprl~\cite{constable1986nuprl}, MetaPRL~\cite{hickey2003metaprl}, and RedPRL~\cite{angiuli2018redprl} establish the extensional lineage, with Anand and Rahli's Coq formalization of Nuprl's semantics providing the verified core precedent and Hickey and Nogin's MetaPRL compiler providing the HOAS empirical baseline; our work adapts this foundation to autonomous agent synthesis.
\item \textbf{Agents Under a Machine Referee:} Kaashoek and Zeldovich~\cite{kaashoek2026machcsl} report a Rocq-checked proof of the xv6 kernel on Sail RISC-V semantics under human-designed intermediate specs, AlphaProof~\cite{hubert2025alphaproof} proves competition mathematics in Lean, and SaltBench (arXiv:2609.11076)~\cite{hickey2026saltbench} measures how an external referee alters agent behavior and supplies the cost bounds cited in Section 1.2; our paper defines the organizational engineering layer these results enable.
\item \textbf{Formal Methods in Industry:} TLA+ at Amazon~\cite{newcombe2015aws}, seL4~\cite{klein2009sel4}, IronFleet~\cite{hawblitzel2015ironfleet}, and Project Everest~\cite{bhargavan2017everest} demonstrate specialist-built verified systems; Pillar 5 examines when agents serve as the verification specialists and generalist SWEs write only specifications.
\item \textbf{Memory and Ownership:} Perceus reuse~\cite{reinking2021perceus,lorenzen2023fp2}, Tofte and Talpin's regions~\cite{tofte1997region}, Iris~\cite{jung2018iris}, and fractional permissions~\cite{boyland2003fractional,ohearn2007resources} provide the mechanisms assembled in Sections 2.4 and 3.2 under HOAS binders in an agent-synthesized IR.
\end{itemize}

\subsection*{7.4 Separation of Existing Baselines and Proposed Components}
This paper is a proposal. To allow readers to weigh each architectural pillar by its current empirical grounding, the table below separates built and measured systems from proposed components.

\begin{center}\small
\begin{tabularx}{\linewidth}{@{}LLL@{}}
\toprule
\textbf{Component} & \textbf{Status} & \textbf{Evidence Base} \tabularnewline
\midrule
Spec-first review gates (lint, vacuity, mutation, non-author reads) & Built and in daily use & Public salt (Lean 4) repository CI~\cite{salt2026} \tabularnewline
Agent evaluation under a machine referee (200-condition pilot) & Built and published & SaltBench (arXiv:2609.11076) \tabularnewline
Mechanized x86-64 user-level ISA semantics & Built and validated & Public x86lean (Lean 4) repository~\cite{x86lean2026} \tabularnewline
HOAS formal compiler construction in CTT & Prototyped in MetaPRL (toolchain not current) & Hickey and Nogin (484 vs. 12,000 lines) \tabularnewline
Verification IR (linear functional SSA, regions, permissions) & Proposed & Defined in this paper \tabularnewline
Thin verified lowering and register allocation via \ensuremath{\alpha}-conversion & Proposed & Targets x86lean semantics \tabularnewline
Verified certificate checker for VIR logic & Proposed & Precedent in Anand and Rahli (ITP 2014) \tabularnewline
Two-tier fleet, semantic merge conflicts, TCB debt ledger & Proposed & Defined in this paper \tabularnewline
Spec-readability experiment and review-surface ratio & Commissioned / protocol frozen before the first reader & Forthcoming empirical study \tabularnewline
\bottomrule
\end{tabularx}
\end{center}

Pillars 1 and 2 rest on the first two rows (salt and SaltBench), Pillar 3 on the third (x86lean), Pillar 4 on the fourth (Hickey and Nogin) and the planned readability experiment, and Pillar 5 is an architectural argument.

\section*{Appendix: Formal CTT Specifications and HOAS VIR Rewrite Rules}
This appendix presents the concrete mathematical definitions and bidirectional term-rewriting rules underpinning Case Study 1 (Section 5), following the HOAS and second-order substitution conventions of Computational Type Theory (Hickey and Nogin) where syntax nodes take the canonical form opname[params]\{v\ensuremath{_{1}}. t\ensuremath{_{1}}; ...; v\ensuremath{_{m}}. t\ensuremath{_{m}}\} and second-order metavariable e[v] performs automatic capture-avoiding substitution e[t].

\subsection*{A.1 Shared Domain Ontology and Extensional Subset Types (CTT)}
Because subset membership proofs in CTT \{ x : A \ensuremath{\mid} P(x) \} are semantic judgments over PERs rather than syntactic pair components, runtime VIR terms carry zero proof-term overhead:

\begin{lstlisting}
(* Base Types and Wire Layout Constants *)
Byte       `\uc{$\equiv$}` { i : Int `\uc{$\mid$}` 0 `\uc{$\le$}` i < 256 }
U32        `\uc{$\equiv$}` { i : Int `\uc{$\mid$}` 0 `\uc{$\le$}` i < 2`\uc[2]{$^{32}$}` }
WireBuf    `\uc{$\equiv$}` Array(Byte)
Principal  `\uc{$\equiv$}` U32
TopicId    `\uc{$\equiv$}` U32
MAGIC      `\uc{$\equiv$}` 0x4D534731  (* "MSG1" wire header magic *)

(* Extensional Subset Type for Valid Network Frames *)
ValidFrame `\uc{$\equiv$}` { b : WireBuf `\uc{$\mid$}` len(b) `\uc{$\ge$}` 24 `\uc{$\wedge$}` read_u32(b, 0) = MAGIC }

(* Semantic Projections over ValidFrame (No Bounds Witnesses Required) *)
frame_sender(b)  `\uc{$\equiv$}` read_u32(b, 4)
frame_topic(b)   `\uc{$\equiv$}` read_u32(b, 8)
frame_pay_off(b) `\uc{$\equiv$}` read_u32(b, 12)
frame_pay_len(b) `\uc{$\equiv$}` read_u32(b, 16)
\end{lstlisting}

\subsection*{A.2 Linear Functional SSA VIR Grammar and Foundational Rewrite Rules}
The VIR is an A-Normal Form (Functional SSA) language where every intermediate operation is bound by an untyped HOAS binder and let\_lin enforces single-use linear ownership:

\begin{lstlisting}
(* Atoms a and Linear Functional SSA Expressions e *)
a ::= i | true | false | v | a`\uc{$_{1}$}` binop a`\uc{$_{2}$}` | a`\uc{$_{1}$}` relop a`\uc{$_{2}$}`

e ::= let_atom{a; v. e[v]}                    (* Pure atom binding *)
    | let_lin{op(a`\uc{$_{1}$}`, ..., a`\uc{$_{n}$}`); b. e[b]}       (* Linear buffer/state binding *)
    | advance{b; offset; s. e[s]}             (* Zero-copy slice advance *)
    | read_u32{b; offset; v. e[v]}            (* Bounded 32-bit wire read *)
    | guard{a_cond; e_ok; e_err}              (* Branching bounds/ACL guard *)
    | return{a} | drop{code}                  (* Terminal outcomes *)
\end{lstlisting}
Capture-avoiding inlining, dead-code elimination, constant folding, and branch pruning require four bidirectional equivalence rules (redex \ensuremath{\leftrightarrow} contractum).

\begin{center}\small
\begin{tabularx}{\linewidth}{@{}LLL@{}}
\toprule
\textbf{Rule Name} & \textbf{Bidirectional HOAS Rewrite Rule} & \textbf{Side-Condition and Role} \tabularnewline
\midrule
{}[beta\_lin] & let\_lin\{v\ensuremath{_{1}}; v\ensuremath{_{2}}. e[v\ensuremath{_{2}}]\} \ensuremath{\leftrightarrow} e[v\ensuremath{_{1}}] & Meta-level capture-avoiding substitution for cross-module inlining \tabularnewline
{}[dce\_atom] & let\_atom\{a; v. e[]\} \ensuremath{\leftrightarrow} e[] & Second-order pattern e[] ensures v \ensuremath{\notin} FV(e) for non-trapping a \tabularnewline
{}[const\_fold] & (i binop j) \ensuremath{\leftrightarrow} \ensuremath{[\![}binop\ensuremath{]\!]}(i, j) & Evaluates constant 32/64-bit arithmetic and guards at synthesis time \tabularnewline
{}[guard\_true] & guard\{true; e\ensuremath{_{1}}; e\ensuremath{_{2}}\} \ensuremath{\leftrightarrow} e\ensuremath{_{1}} & Prunes unreachable error branches once composite guard holds \tabularnewline
\bottomrule
\end{tabularx}
\end{center}

\subsection*{A.3 Step-by-Step Derivation: Fusing the Message Router Pipeline}
To fuse Wire Header Parsing (Spec A), Token-Bucket Rate Limiting (Spec C), and Security ACL Verification (Spec B) into a single zero-copy pass, the agent applies three domain-specific algebraic rewrites over buffer slices:

\begin{lstlisting}[basicstyle=\ttfamily\footnotesize]
(* Rule 1: Slice-Advance Coalescing *)
[slice_fuse]   advance{b; i; s`\uc{$_{1}$}`. advance{s`\uc{$_{1}$}`; j; s`\uc{$_{2}$}`. e[s`\uc{$_{2}$}`]}}
           `\uc{$\leftrightarrow$}`   advance{b; i + j; s`\uc{$_{2}$}`. e[s`\uc{$_{2}$}`]}      (when i `\uc{$\ge$}` 0 `\uc{$\wedge$}` j `\uc{$\ge$}` 0 `\uc{$\wedge$}` i + j `\uc{$\le$}` len(b))

(* Rule 2: Guard Hoisting & Conjunction Merging *)
[guard_merge]  guard{c`\uc{$_{1}$}`; guard{c`\uc{$_{2}$}`; e_ok; e_err}; e_err}
           `\uc{$\leftrightarrow$}`   guard{c`\uc{$_{1}$}` `\uc{$\wedge$}` c`\uc{$_{2}$}`; e_ok; e_err}

(* Rule 3: Offset Read Commutation Across Disjoint State *)
[read_comm]    read_u32{b; i; v`\uc{$_{1}$}`. let_lin{upd_bucket(st, v`\uc{$_{1}$}`); st'. read_u32{b; j; v`\uc{$_{2}$}`. e[v`\uc{$_{1}$}`; st'; v`\uc{$_{2}$}`]}}}
           `\uc{$\leftrightarrow$}`   read_u32{b; i; v`\uc{$_{1}$}`. read_u32{b; j; v`\uc{$_{2}$}`. let_lin{upd_bucket(st, v`\uc{$_{1}$}`); st'. e[v`\uc{$_{1}$}`; st'; v`\uc{$_{2}$}`]}}}
\end{lstlisting}
Applying [beta\_lin], [guard\_merge], [slice\_fuse], and [dce\_atom] in sequence yields the fused single-pass, zero-allocation VIR term:

\begin{lstlisting}
(* Synthesized & Fused Single-Pass VIR Term *)
guard{ len(b`\uc{$_{0}$}`) `\uc{$\ge$}` 24;
  read_u32{ b`\uc{$_{0}$}`; 0; magic.
  read_u32{ b`\uc{$_{0}$}`; 4; sender.
  read_u32{ b`\uc{$_{0}$}`; 8; topic.
  guard{ magic = MAGIC `\uc{$\wedge$}` can_access(sender, topic) `\uc{$\wedge$}` has_tokens(bucket_st, sender);
    let_lin{ dec_tokens(bucket_st, sender); bucket_st'.
    advance{ b`\uc{$_{0}$}`; 24; payload_slice.
      return{ (bucket_st', payload_slice) } } };
    drop{ ERR_UNAUTHORIZED_OR_RATE_LIMITED } } } } };
  drop{ ERR_TRUNCATED_FRAME } }
\end{lstlisting}

\subsection*{A.4 Lowering to Scoped Assembly and Register Allocation via \ensuremath{\alpha}-Conversion}
Following Section 4 of Hickey and Nogin, the target instruction set is formalized using scoped immutable register bindings where instructions modifying a destination register bind a fresh meta-variable for the continuation (ADD o\ensuremath{_{1}}, \%v\ensuremath{_{2}}, \ensuremath{\lambda}v\ensuremath{_{3}}. e[v\ensuremath{_{3}}]), memory writes use non-binding forms (MOV \%v\ensuremath{_{1}}, offset(\%v\ensuremath{_{2}}); e), and register pressure spills use scoped spill instructions (SPILL \%v\ensuremath{_{1}}, \ensuremath{\lambda}s. e[spill[v\ensuremath{_{1}}, s]]). An untrusted Chaitin-style graph-coloring heuristic proposes a mapping from VIR variables (b\ensuremath{_{0}}, magic, sender, topic, payload\_slice) to physical x86-64 registers (\%rdi, \%eax, \%ecx, \%edx, \%rsi), which the proof kernel verifies purely as a capture-avoiding \ensuremath{\alpha}-conversion:

\begin{lstlisting}[basicstyle=\ttfamily\footnotesize]
(* Exhibit A.4: Lowered Scoped x86-64 Assembly for the Fused Router Fast-Path *)
(* Input: %rdi = b`\uc{$_{0}$}` base ptr, %rsi = b`\uc{$_{0}$}` byte len, %rdx = bucket_st ptr *)
CMP   $24, %rsi
JB    then drop_truncated else      (* unsigned: the length is a byte count *)
MOV   0(%rdi), `\uc{$\lambda$}`%eax.                  (* magic  `\uc{$\leftarrow$}` b`\uc{$_{0}$}`[0..3]  *)
CMP   $0x4D534731, %eax
JNE   then drop_unauthorized else
MOV   4(%rdi), `\uc{$\lambda$}`%ecx.                  (* sender `\uc{$\leftarrow$}` b`\uc{$_{0}$}`[4..7]  *)
MOV   8(%rdi), `\uc{$\lambda$}`%r8d.                  (* topic  `\uc{$\leftarrow$}` b`\uc{$_{0}$}`[8..11] *)
LEA   ACL_BITMAP(%rip), `\uc{$\lambda$}`%r9.        (* RIP-relative takes no index *)
BT    %r8, 0(%r9, %rcx, 8)            (* constant-time ACL bit test *)
JNC   then drop_unauthorized else
CMP   $0, BUCKET_ARR(%rdx, %rcx, 4)   (* the bucket is tested before it is written *)
JLE   then drop_rate_limited else
SUB   $1, BUCKET_ARR(%rdx, %rcx, 4)   (* in-place linear token update *)
LEA   24(%rdi), `\uc{$\lambda$}`%rax.                 (* payload_slice `\uc{$\leftarrow$}` b`\uc{$_{0}$}` + 24 *)
SUB   $24, %rsi                       (* payload length = len - 24 *)
JMP   emit_slice                      (* on entry: %rax = payload, %rsi = its length *)
\end{lstlisting}
At lowering time, the functional linear state token bucket\_st erases into the exclusive base-pointer capability \%rdx, so that let\_lin\{dec\_tokens(bucket\_st, sender); bucket\_st'. ...\} lowers to the in-place memory instruction SUB \$1, BUCKET\_ARR(\%rdx, \%rcx, 4), placed after the bucket is tested, with the processor condition flags (JB, JNE, JNC, JLE) modeled as scoped branching continuations.

\textbf{What is checked.} Exhibit A.4 is the lowering as repaired. Its machine code, assembled by a standard assembler, is checked in the x86lean semantics~\cite{x86lean2026} against the router's functional specification: for every start state whose registers and memory encode the specification's inputs (the buffer, the ACL table and the bucket state), in which topic $<$ 64 and the buffer does not overlap the sender's bucket cell (the one cell written), the program stops at an exit for the specification's outcome (the combined unauthorized-or-rate-limited outcome may take either of its two labels), with memory unchanged on each drop, and on emit\_slice with \%rax $=$ $b_0 + 24$, \%rsi $=$ $\mathit{len} - 24$ and exactly one bucket cell decremented (theorem repaired\_refines in Router/Refines.lean at x86lean commit adbf22f995d2~\cite{o95spike2026}). A first draft of this exhibit was not a refinement, and two concrete states refute it: it compared the length as a signed number (a length of $2^{63}$ or more was dropped as truncated), and it decremented the bucket before testing it (a bucket of 0 became 0xFFFFFFFF on a drop). It also used a RIP-relative address with an index register, which cannot be encoded, and its exit passed no slice length. The listing above repairs those four. The fifth, topic $<$ 64, is a precondition of the theorem rather than a repair, since the 32-bit TopicId of A.1 does not supply it. \textbf{Not claimed:} the $\alpha$-conversion check and the register allocator described above are not built; the theorem is about these instructions, not about a procedure that produces them; and it relates one run of the program to one application of the specification. A second theorem covers two successive calls with the bucket state threaded through: the second call refines the specification applied to the state the first call returned, so two emits from one sender need at least two tokens at the start and leave that sender's bucket cell two lower (two\_calls in Router/MultiCall.lean at the same commit~\cite{o95spike2026}, and two\_emits\_same\_sender in compose/Compose/RateLimit.lean at vir commit 7595dfd8d128~\cite{vir2026}). Its limits: it is about two calls, not $n$; between them the caller writes only the second call's buffer; and since the lowering does not bound the sender, that buffer must miss a $2^{35}$-byte ACL table and a $2^{34}$-byte bucket array, and the two tables must not overlap, which requires about 48 GiB of address space for the tables. That last limit is a property of the object as written, not of the proof. The theorem holds in the x86lean model at a pinned public revision; this program has not been run on hardware, and the model's differential validation covers instruction forms, not this program. Code is not in memory in the model, so self-modifying code is outside what is checked.


\end{document}